\documentclass[a4paper,11pt]{article}
\usepackage{pos}
\usepackage{physics}
\usepackage{graphicx}

\title{Continuum limit of two-dimensional multiflavor scalar gauge theories}
\author*[a]{Alessio Franchi}
\author[a]{Claudio Bonati}
\author[b]{Andrea Pelissetto}
\author[a]{Ettore Vicari}

\affiliation[a]{Dipartimento di Fisica, Università di Pisa, and INFN Sezione di Pisa,\\
Largo Bruno Pontecorvo 3, Pisa, Italy}

\affiliation[b]{Dipartimento di Fisica, Università di Roma “La Sapienza”, and INFN Sezione di Roma,
P.le Aldo Moro 2, Roma, Italy.}

\emailAdd{alessio.franchi@phd.unipi.it}
\emailAdd{claudio.bonati@unipi.it}
\emailAdd{andrea.pelissetto@roma1.infn.it}
\emailAdd{ettore.vicari@unipi.it}

\abstract{We address the interplay between local and global symmetries by analyzing the continuum limit of two-dimensional multicomponent scalar lattice gauge theories, endowed by non-Abelian local and global invariance. These theories are asymptotically free. By exploiting Monte Carlo simulations and finite-size 
scaling techniques, we provide numerical results concerning the universal
behavior of such models in the critical regime. Our results support the
conjecture that two-dimensional multiflavor scalar models have the same
continuum limit as the $\sigma$-models 
associated with symmetric spaces that have the same global symmetry.}

\FullConference{%
 The 38th International Symposium on Lattice Field Theory, LATTICE2021
  26th-30th July, 2021
  Zoom/Gather@Massachusetts Institute of Technology
}

\begin{document}
\maketitle

\section{Introduction}
Local and global symmetries play a crucial role in determining the main
features of a many-body physical system. Non-Abelian gauge invariance is often
associated with color confinement and asymptotic freedom \cite{Wilson-74},
while global invariance and the corresponding symmetry-breaking pattern is related to phase transitions and universality. We focus on the intertwining of these two symmetries in the presence of a critical phenomenon, to better understand the role played by local (color) and global (flavor) degrees of freedom in determining the modes that become critical and therefore the universality class observed approaching the continuum limit. Due to the Mermin-Wagner theorem \cite{MW-66},  two-dimensional systems cannot undergo 
a spontaneous breaking of a continuous symmetry at finite temperature. However,
a critical behavior may be expected in the asymptotic zero-temperature limit,
as it happens in 2D non-linear $\sigma$ models (NLSMs) \cite{ZJ-book}. Indeed, these theories develop an exponentially diverging correlation length in the low-temperature regime
\begin{equation}
    \xi \underset{T\to 0^+}{\sim}T^{-p} e^{c/T}
\end{equation}
similarly to QCD, the theory of strong interactions. Generalizations of NLSMs,
e.g. gauged $\sigma$-models defined on a symmetric space $G/H$, are 
expected to be asymptotically free as well \cite{BHZ-80}.

In this proceeding, we review some results  on Multiflavor Scalar Gauge (MSG) 
theories on a two-dimensional lattice, characterized by a 
non-Abelian SO($N_c$) gauge symmetry and O($N_f$) global invariance,
as paradigmatic examples to discuss the critical properties of
MSG models \cite{BFPV-20-ong}.  We only
consider fields transforming according to the fundamental representation of the
gauge group.  To investigate the critical
behavior of these lattice systems, we performed finite-temperature Monte 
Carlo simulations
and a detailed analysis of the minimum-energy configurations.
All results support the following conjecture: a two-dimensional MSG model has
 the same continuum limit as the NLSM defined on a symmetric space, possessing
the same global symmetry as the gauge model. This is confirmed by the 
study of the simplest formulation, i.e., in the absence of any local potential 
in the Hamiltonian. In this case 
$\mathbb{R}\text{P}^{N_f-1}$ critical behavior is observed for any $N_c\geq3$, 
$N_c$ and $N_f$ being the number of colors and flavors, respectively.

However, different symmetry-breaking patterns and critical behaviors may be
driven by the introduction of a quartic potential that does not
break explicitly any symmetry of the lattice model.
Indeed, by varying the potential parameters, one can vary the structure of the
scalar-field vacuum, which crucially determines  the continuum limit 
realized for $\beta\to \infty$.
For multiflavor scalar theories in the unit-length limit, we considered 
the most general gauge-invariant quartic potential that preserves the O($N_f$) 
global invariance of the model, to understand the role that the potential
plays in determining the universality classes and 
continuum limits of MSG models. 
We present here some numerical results for $N_c=3$ with a nonvanishing 
quartic coupling [they are taken from \cite{BFPV-21-Adjoint_2d}, that 
studied MSGs with fields in the adjoint representation of SU(2), which
is equivalent to the fundamental representation of SO(3)].
These results can be extended to more general models undergoing transitions
belonging to more exotic universality classes, associated with a NLSM defined 
on a symmetric space. In particular, for $N_f>N_c$ and appropriate quartic
couplings, the
continuum limit is described by a NLSM defined on the 
quotient group $\text{SO}(N_f)/(\text{SO}(N_c)\otimes\text{SO}(N_f-N_c))$
\cite{BHZ-80, BF-21-symmetric_spaces}, corroborating our general conjecture.

\section{Lattice model and observables}
We define the model on a square lattice with periodic boundary conditions. On
each site $x$ of the lattice, we define a real matrix $\phi^{if}_x$ (where
$i=1,..,N_c$ and $f=1,..,N_f$ are the color and flavor indices, respectively)
satisfying the unit-length constraint
\begin{equation}
    \Tr \phi^t_x \phi_x = 1\,.
\end{equation}
To implement the SO($N_c$) gauge symmetry, we introduce link variables $U_{x,\mu}\in \text{SO}(N_c)$ on each link connecting two nearest-neighbor sites, according to the Wilson prescription. 
The Hamiltonian and the partition function of the lattice model read as follows
\begin{align}
    H_{\text{MSG}} = -N_f\sum_{x, \mu} \Tr \phi^t_x U_{x, \mu} &\phi_{x+\mu} - \frac{\gamma}{N_c}\sum_{x} \Tr U_{x, \mu}U_{x+\mu, \nu}U^t_{x+\nu, \mu}U^t_{x, \nu}+V(\phi_x)\\
    Z &= \sum_{\{\phi, U\}}e^{-\beta H}\,, \quad \beta = 1/T\,,
\end{align}
where V($\phi_x$), as discussed in the introduction, is a gauge-invariant
potential term which preserves the symmetries of the lattice model. We 
consider the most general quartic potential compatible with this requirement, 
i.e., 
\begin{equation}
    V(\phi_x) = w \sum_x \Tr \phi_x^t \phi_x \phi^t_x \phi_x\,.
    \label{quartic_interaction}
\end{equation}
The system is characterized by SO($N_c$) gauge invariance
\begin{equation}
    \phi_x \mapsto \phi'_x=W_x \phi_x,\quad \quad U_{x,\mu} \mapsto U'_{x,\mu}=W_x U_{x,\mu} W^t_{x+\mu}, \quad \quad W_x \in \text{SO}(N_c)\,, 
\end{equation}
and O($N_f$) global invariance
\begin{equation}
    \phi_x \mapsto \phi'_x = \phi_x M, \quad \quad U_{x,\mu} \mapsto U'_{x,\mu}=U_{x,\mu}, \quad \quad M\in \text{O}(N_f).
\end{equation}
To determine the critical behavior associated with the breaking of the
SO($N_f$) global symmetry, we study 
the condensation of a traceless spin-2 symmetric operator $Q_x$
\begin{equation}
    B^{fg}_x=\sum_{i=1}^{N_c} \phi^{if}_x\phi^{ig}_x \quad \quad Q^{fg}_x = B^{fg}_x - \frac{\delta^{fg}}{N_f}\,,
\end{equation}
which is the simplest local gauge-invariant operator that can be defined 
on the lattice.
In particular, to classify the models at criticality, we focus on two
Renormalization Group (RG) invariant quantities such as the quartic Binder
cumulant $U$ and the ratio $R_\xi=\xi/L$, 
\begin{equation}
    \xi^2 = \frac{1}{4 \sin^2{\frac{\pi}{L}}}\frac{\Tilde{G}(0) - \Tilde{G}(p_m)}{\Tilde{G}(p_m)},\quad U = \frac{\expval{\mu_2^2}}{\expval{\mu_2}^2},\quad \mu_2 = \frac{1}{V^2}\sum_{x, y} \Tr Q_x Q_y\,,
\end{equation}
$\xi$ being the second-moment correlation length, $L$ the lattice size,
$p_m$ the minimum value of the momentum consistent with periodic
boundary conditions, and $\Tilde{G}$ the Fourier transform of the two-point
function $G(x-y)=\expval{\Tr Q_x Q_y}$. See \cite{BFPV-20-ong} for additional
technical details. Given two RG invariant quantities, for instance $U$ and
$R_\xi$, we determine $U(R_\xi)$, i.e., how the Binder parameter 
$U$ depends on $R_\xi$, for several lattice sizes. According to the
Finite-Size Scaling (FSS) theory, if two models belong to the 
 same universality class, in the FSS limit $U(R_\xi)$ is the same, i.e.,
the function $F(R_\xi)$ defined by
\begin{equation}
    U(R_\xi) \underset{L \to +\infty}{\approx} F(R_\xi)\,
    \label{Binder_strategy}
\end{equation}
is the same.
For finite-size systems Eq.~(\ref{Binder_strategy}) 
 holds apart from scaling corrections, whose leading behavior is controlled by 
the RG scaling dimension of the lowest-dimensional irrelevant operator appearing in the theory. 
In the case of asymptotically free models, such as two-dimensional vector models, corrections decrease as $L^{-2}$, multiplied by powers of $\ln L$ \cite{CP-corrections}.
This strategy allows us to compare different models at criticality without 
tuning any non-universal parameter, as the function $F(R_\xi)$ only depends on the boundary conditions and the universality class associated with the system.

\section{Minimum-energy configurations}

The study of the minimum-energy configurations is a very useful and handy way to predict the critical behavior of two-dimensional multiflavor scalar systems. 
Practically, we perform simulations on very small systems for very large values
of $\beta$, and then we extrapolate the expectation values, to 
obtain the  $\beta\to+\infty$ limit. Here, 
we restrict the discussion to $\Tr B^2_x$, whose average value 
$\expval{\Tr B^2_x}$ allows us to determine the structure of the 
scalar-field vacuum. In agreement with general considerations, 
different results, corresponding the different vacuum structures,
are obtained for $w > 0$ and $w \le 0$, see, for instance,
Table~\ref{table_minima}.
\begin{table}[!t]
    \centering
    \begin{tabular}{c|c|c}
         \hline \hline
         $(N_c, N_f)$&$\expval{\Tr B^2_x}_{w=0}$&$\expval{\Tr B^2_x}_{w=1}$  \\
         \hline 
         (3, 2)&0.9998(4)&0.49981(13) \\
         (3, 3)&0.9999(5)&0.3331(2) \\
         (3, 4)&1.0000(4)&0.3332(2) \\
         (4, 3)&1.00000(1)& - \\
         (4, 4)&1.00001(2)& -  \\
         \hline \hline
    \end{tabular}
    \caption{Estimates of $\expval{\Tr B^2_x}$ on 
minimum-energy configurations for square lattices of size $L=4$ and several color-flavor combinations.}
    \label{table_minima}
\end{table}

These results can be easily understood.
Applying the singular value decomposition to $\phi^{ia}_x$, one can prove that 
two minima exists and, correspondingly, $\Tr B^2$ is either equal to 1 or 
to $1/q$, where $q=\min[N_c, N_f]$. More precisely, in the limit
$\beta\to+\infty$, the rectangular matrix $\phi_x$ can be casted in one 
of the two following forms, depending on the quartic coupling sign:
\begin{equation}
    \begin{aligned}
    w \leq 0:& \ \phi^{ia}_x = s^i_x z^a_x, \quad \quad \text{where} \quad s^i_x, z^a_x \ \text{are unit-length vectors}\\
    w > 0:& \ \phi^{ia}_x = \sqrt{\frac{1}{q}}\sum^q_{k=1}C_x^{ik}O_x^{ak}, \quad \ \text{where} \quad \text{$C\in$ O($N_c$), $O\in$ O($N_f$).}
    \end{aligned}
    \label{cases_different_quartic_coupling}
\end{equation}
For $w\leq0$, $\Tr B^2_x$ is $1$ (see App.~A of 
\cite{BFPV-21-Adjoint_2d} for a discussion of the case $w=0$)
and  the bilinear operator $B^2_x$ reduces to a projector 
$P^{fg}_x$ onto a one-dimensional space
\begin{equation}
    B^{fg}_x = z^f_x z^g_x = P^{fg}_x\,, \quad \text{with} \quad P^2_x=P_x\,.
\end{equation}
If we assume that the dynamics in the gauge model is completely determined by the fluctuations of the order parameter $B_x$, by substituting 
Eq.~\eqref{cases_different_quartic_coupling} in the Hamiltonian, 
we identify the effective scalar model as the $\mathbb{R}\text{P}^{N_f-1}$ model \cite{BFPV-20-ong, BFPV-21-Adjoint_2d}. Indeed, the 
standard nearest-neighbor $\mathbb{R}\text{P}^{N_f-1}$ action is given by
\begin{equation}
    \text{H}_{R\text{P}^{N-1}} = - J \sum_{x, \mu} \Tr P_x P_{x+\mu}, \quad P^{fg}_x = \phi^f_x\phi^g_x
\end{equation}
This is fully confirmed by the results of Table \ref{table_minima} 
and by the FSS analyses we will provide in the next section.

For $w > 0$, the scalar vacuum is different and 
$\expval{\Tr B^2_x}$ is equal to $1/q$, 
as verified in Table~\ref{table_minima}. 
We will not proceed further discussing this case, as the topic is more technical and out of the scope of this proceeding. We only mention that 
preliminary results support a critical behavior associated with a NLSM 
defined on the symmetric space
$\text{SO}(N_f)/(\text{SO}(N_c)\otimes\text{SO}(N_f-N_c))$, if $N_f>N_c$
\cite{BF-21-symmetric_spaces}. Note that these universality classes depend
on the number of colors $N_c$---at variance with the case  
$w\leq0$---and have a peculiar symmetry under $N_c \mapsto N_f - N_c$.  

\section{Numerical results}

\begin{figure}[!t]
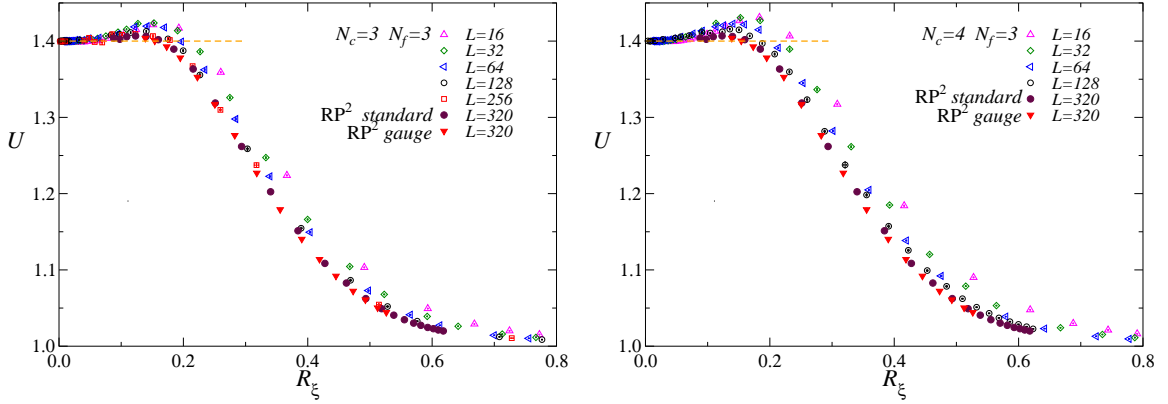

    \centering
    \includegraphics[width=0.49\textwidth]{urxi_nc3_nf3.eps}\hfill
    \includegraphics[width=0.49\textwidth]{urxi_nc4_nf3.eps}
    \caption{Left: Plot of $U$ versus $R_\xi$ for $N_f=3$ and $w=0$.
Left: results for $N_c=3$; right: results for $N_c=4$.
In both plots
MSG data are compared with the scaling curve of two different formulation of the $\mathbb{R}\text{P}^2$ model (see \cite{BFPV-20} for more details).}
    \label{figures_urxi_3flav}
\end{figure}

We present some FSS analyses taken from 
\cite{BFPV-20-ong, BFPV-21-Adjoint_2d}. We will make extensive use of the 
relation Eq.~\eqref{Binder_strategy}, comparing plots of the Binder 
cumulant $U$ as a functions of $R_\xi$: if two theories have
the same critical behavior, the asymptotic curve $F_U(R_\xi)$ is 
the same in the FSS limit. 
We first show data in the absence of a potential term ($w=0$) for
 $N_f=3$ and $N_c=3,4$, see the left and right panel of 
Fig.~\ref{figures_urxi_3flav}. In both cases, increasing the lattice size, MSG data approach monotonically the asymptotic curve of the $\mathbb{R}\text{P}^2$ model: tiny finite-size deviations are interpreted as scaling corrections. 
The figure provides numerical evidence that MSG theories with $N_f=3$ and 
non-Abelian gauge symmetry (without any potential term in the Hamiltonian) 
are in the same universality class as the $\mathbb{R}\text{P}^2$ vector model.

\begin{figure}[!t]
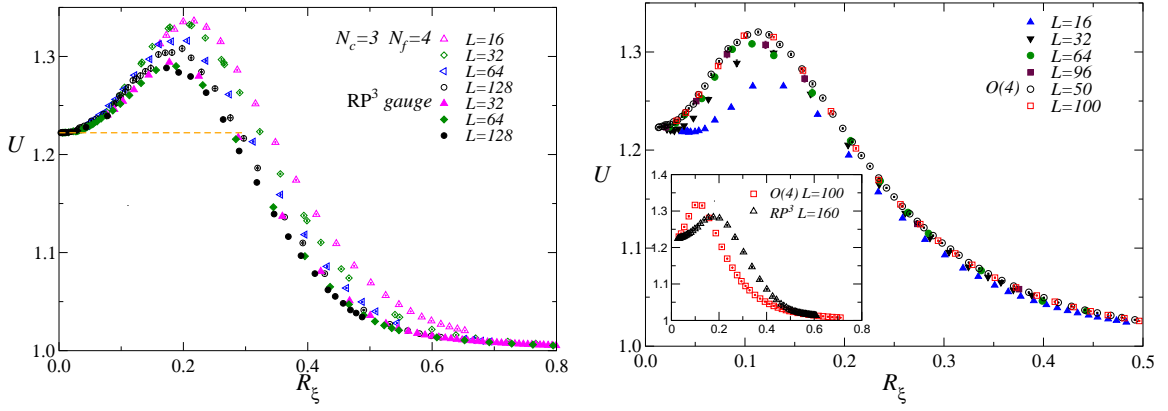

    \centering
    \includegraphics[width=0.49\textwidth]{urxi_nc3_nf4.eps}\hfill
    \includegraphics[width=0.49\textwidth]{urxi_nc2_nf4_v10_with_rp3.eps}
    \caption{Plot of $U$ versus $R_\xi$ for $N_c=3, N_f=4$. 
In the left panel we show results for $w=0$, in the right panel results for 
$w = 10$. 
MSG data are compared with the scaling curve of the 
$\mathbb{R}\text{P}^3$ model (left) and of the O(4) vector model (right).
The scaling curves of these two models are also reported in the inset
(right panel).
}
    \label{figures_urxi_4flav_and_potential}
\end{figure}

We then present results for $N_c=3, N_f=4$ (note that $N_f>N_c$) and different
values of the quartic coupling, $w=0$ and 10, see 
Fig.~\ref{figures_urxi_4flav_and_potential}. In the absence of a potential term (left panel), we still observe a $\mathbb{R}\text{P}^{N_f-1}$ critical behavior: 
the four-flavor scaling curve is consistent with $\mathbb{R}\text{P}^3$ data, 
within scaling corrections. This is in agreement with the analysis 
of the minimum energy configurations: for $w=0$ one obtains
 $\expval{\Tr B^2_x}=1$ as in the $\mathbb{R}\text{P}^3$ model,
see Table \ref{table_minima}.
A different continuum limit is observed for $w>0$. 
In the right panel of Fig.~\ref{figures_urxi_4flav_and_potential}, 
we present results for $w=10$, which are in complete agreement with an 
O(4) critical behavior. As shown in the inset, $\mathbb{R}\text{P}^3$ and O(4) 
data for the largest sizes at our disposal (they provide very good 
approximations of the universal curves $F_U(R_\xi)$ associated with each of the two universality classes) are clearly different, so the two different continuum limits can be easily distinguished. 

\section{Conclusions}

We presented numerical studies of multiflavor scalar models with 
a non-Abelian
SO($N_c$) gauge symmetry, an O($N_f$) global invariance and 
fields transforming in the fundamental representation of the gauge group.
The continuum limit that is realized in the $\beta\to+\infty$ limit crucially depends on the structure of the scalar field vacuum. In the absence of a quartic potential term (actually if $w\leq0$, see \cite{BFPV-21-Adjoint_2d}), the critical behavior is expected to be associated with a projective-space
$\sigma$ model. This conjecture is confirmed by 
the analyses of the minimum-energy configurations. 
We obtained $\expval{\Tr B^2_x}=1$, which implies that 
the lattice model is effectively
described by a theory of projectors. We also performed Monte Carlo simulations
for $(N_c=3,4$ and $N_f=3)$ and $(N_c=3, N_f=4)$. All results
are in agreement with a critical behavior belonging to the
 $\mathbb{R}\text{P}^{N_f-1}$ universality class. 
These findings can be extended in several directions, some of which have been
already explored in the literature. 
One can consider complex scalar fields transforming in the fundamental
representation of the SU($N_c$) gauge group with a U($N_f$) global symmetry
\cite{BPV-20-qcd2}. Analogous arguments to the ones presented in this
proceeding support in this case a $\mathbb{C}\text{P}^{N_f-1}$ critical
behavior. Alternatively, one may consider other group representations.
Specifically in \cite{BFPV-21-Adjoint_2d}, for instance, the authors considered
scalar fields transforming in the adjoint representation of the SU($N_c$) gauge
group. A different possibility is to consider the effects of a positive
quartic coupling in driving more complex critical behaviors, whenever
$N_f>N_c$. As we already mentioned, in that case we expect the
universality class to be associated with NLSMs defined 
on the symmetric spaces
$\text{SO}(N_f)/(\text{SO}(N_c)\otimes\text{SO}(N_f-N_c))$ \cite{BHZ-80,
BF-21-symmetric_spaces}. Preliminary results, as well as all numerical
data we gathered, seem to strengthen the conjecture according to which MSG 
theories have the same continuum limit as NLSMs defined on
a symmetric spaces possessing the same global symmetries as the lattice model.

\end{document}